\documentclass[useAMS]{mn2e}
\usepackage{graphicx}
\usepackage[authoryear]{natbib}
\usepackage{times}

\newcommand{\ulas}{{ULAS\,J1120+0641}}

\title[X-rays from z=7.1]
{X-rays from the redshift 7.1 quasar ULAS~J1120+0641}

\author[M.J. Page et al.]
{M.J. Page$^{1}$, 
C. Simpson$^{2}$, 
D.J. Mortlock$^{3,4}$,
S.J. Warren$^{3}$,
P.C. Hewett$^{5}$, 
\and
B.P. Venemans$^{6}$ and
R.G. McMahon$^{5}$
\\ 
\\
$^{1}$Mullard Space Science Laboratory, University College London,
Holmbury St Mary, Dorking, Surrey, RH5 6NT, UK\\
$^{2}$Astrophysics Research Institute, Liverpool John Moores University, 
 Liverpool Science Park, 146 Brownlow Hill, Liverpool L3 5RF, UK\\
$^{3}$Astrophysics Group, Imperial College London, Blackett Laboratory, 
Prince Consort Road, London, SW7 2AZ, UK\\
$^{4}$Department of Mathematics, Imperial College London, London, SW7 2AZ, UK\\
$^{5}$Institute of Astronomy, University of Cambridge, Madingley Road, 
Cambridge CB3 0HA, UK\\
$^{6}$Max-Planck Institute for Astronomy, K\"onigstuhl 17, 69117 Heidelberg, 
Germany\\
}

\begin{document}

\date{Accepted 2014 January 29. Received 2014 January 23; in original form 2013 November 07}

\pagerange{\pageref{firstpage}--\pageref{lastpage}} 
\pubyear{2013}
\maketitle

\label{firstpage}

\begin{abstract}
We present X-ray imaging and spectroscopy of the redshift $z=7.084$ 
radio-quiet
quasar ULAS\,J112001.48+064124.3 obtained with {\em Chandra} and {\em XMM-Newton}. The quasar is detected as a point source with both observatories. The {\em Chandra} observation provides a precise position, confirming the association of the X-ray source and the quasar, while a sufficient number of photons is detected in the {\em XMM-Newton} observation to yield a meaningful X-ray spectrum. In the {\em XMM-Newton} observation the quasar has a 2-10 keV luminosity of $4.7\pm0.9\ \times 10^{44}$~erg~s$^{-1}$ and a spectral slope $\alpha = 1.6^{+0.4}_{-0.3}$ (where $f_{\nu} \propto\nu^{-\alpha}$). 
The quasar appears to have dimmed in the 15 months between the two observations, with a 2-10 keV luminosity of $1.8^{+1.0}_{-0.7}\times 10^{45}$~erg~s$^{-1}$ during the {\em Chandra} observation. We derive optical to X-ray spectral slopes $\alpha_{OX}$ of $1.76\pm0.07$ and $1.54^{+0.09}_{-0.08}$ at the times of the {\em XMM-Newton} and {\em Chandra} observations respectively, consistent with the range of $\alpha_{OX}$ found in other quasars of comparable ultraviolet luminosity.
The very soft X-ray spectrum suggests that the quasar is accreting above the Eddington rate,
$L/L_{\rm Edd} = 5^{+15}_{-4}$, compared to $L/L_{\rm Edd} = 1.2^{+0.6}_{-0.5}$ derived from the rest-frame ultraviolet. 
 Super-Eddington accretion would help to reduce the discrepancy between the age of the quasar implied by the small size of the ionized near zone in which it sits ($<10^{7}$ years), and the characteristic e-folding time ($2.5\times 10^{7}$~years if $L/L_{\rm Edd}=2$). Such super-Eddington accretion would also alleviate the challenging constraints on the seed black hole mass
provided that the quasar has been rapidly accreting throughout its history.
The remnant of an individual population III star is a plausible progenitor if an average $L/L_{\rm Edd}>1.46$ has been maintained over the quasar's lifetime.
\end{abstract}


\section{Introduction}
\label{sec:introduction}
The dark ages of the Universe, that followed recombination, ended when
the ultraviolet (UV) radiation from the first luminous objects reionized
the intergalactic medium. It was during this epoch of reionization, 
which ended at a
redshift $z\sim 6$ \citep{fan06}, that
the first massive black holes blazed as quasars and the first
star-forming galaxies assembled. For quasars in particular, the epoch of
reionization offers insights and constraints into the process by which
massive black holes are formed. An accreting black hole growing at the
Eddington rate with a radiative efficiency of 10 per cent has a mass
e-folding timescale \citep[also known as the Salpeter
time;][]{salpeter64} of $5\times 10^{7}$~years
\footnote{
The e-folding timescale for Eddington-limited accretion is $\tau = \epsilon c \sigma_{\rm T} / (4 \pi G (1 - \epsilon) m_{p})$, where $\epsilon$ is the efficiency and $\sigma_{\rm T}$ is the Thomson cross section.
}. 

For the highest
redshift quasar yet found, ULAS\,J112001.48+064124.3 \citep[hereafter
\ulas;][]{mortlock11}, only 13 such e-folding times have elapsed, corresponding to a factor of 
$4.4\times 10^{5}$
increase in mass, 
between $z=30$, the earliest epoch at which stars are thought to have 
formed \citep{bromm09}, 
and
$z=7.084$, at which the quasar is observed \citep{venemans12}. 
  The mass of the black hole in \ulas\ has been estimated from the
  Mg\,{\sc ii} emission line in two studies, yielding
  $2.0^{+1.5}_{-0.7}\times 10^{9}$~${\rm M}_{\odot}$ \citep{mortlock11} and
  $2.4\pm0.2\ \times 10^{9}$~${\rm M}_{\odot}$ \citep{derosa13} but note that
  in this latter case the uncertainties do not include the systematic
  uncertainty inherent in the estimation method. Both studies find 
  that the bolometric luminosity is consistent with the Eddington 
luminosity: \citet{mortlock11} find $L/L_{\rm Edd} = 1.2^{+0.6}_{-0.5}$, 
while \citet{derosa13} find $L/L_{\rm Edd} = 0.5$ with a factor of 2 
systematic error. 
The deepest upper limit on the radio emission from \ulas\ implies that 
the 1.4~GHz to 4400\AA\ rest-frame flux density ratio is  
$R_{1.4}^{*}<4.3$, classifying the source as radio-quiet \citep{momjian14}.
Assuming a black-hole 
mass of $2.4 \times 10^{9}$~${\rm M}_{\odot}$,
\ulas\ would have required a 
seed black hole
of $5.4\times 10^{3}$~M$_{\odot}$ if it grew under Eddington-limited accretion
at an efficiency of 10 per cent for the entire period. 
Thus \ulas\ offers challenging constraints for the 
progenitor black holes of active galactic nuclei (AGN), 
the rate at which they can grow, or both.

X-ray emission is an ubiquitous property of AGN, emanating from a
corona of the inner accretion disc, within a few hundred Schwarzschild
radii of the massive black hole.  In this paper we present X-ray
observations of \ulas\ and consider their implications for its
accretion rate and growth history. In Section~\ref{sec:observations}
we detail the observations and the data reduction. The results are
presented in Section \ref{sec:results}.  The X-ray emission properties
of the quasar are discussed and conclusions are drawn in Section
\ref{sec:discussion}.
Throughout we adopt the cosmological parameters from the \citet{planck13}:
$H_{0}=67.3$~km~s$^{-1}$, $\Omega_{\Lambda}=0.685$ and $\Omega_{\rm
  m}=0.315$. We define power law spectral indices $\alpha$ such that
$f_{\nu} \propto \nu^{-\alpha}$ (the equivalent form in terms of the
photon index $\Gamma$ is defined as $N(E) \propto E^{-\Gamma}$ where
$\Gamma$ = $\alpha + 1$). X-ray fluxes and luminosities have been
corrected for Galactic absorption equivalent to $N_{H}=5.07\times
10^{20}$~cm$^{-2}$ \citep{kalberla05}.  Unless stated otherwise, all
uncertainties are given at 1\,$\sigma$.

\section{Observations and data reduction}
\label{sec:observations}

\subsection{\em Chandra}
\label{sec:chandraobs}

\ulas\ was observed on 2011 February 4 for 16~ks with {\em Chandra}.
The target fell on the Advanced CCD Imaging Spectrometer
(ACIS) S3 chip, which was operated in full-frame, timed-exposure mode, 
with faint 
telemetry format. 
The data were processed with {\sc ciao} version 4.5
\citep{fruscione06}
and registered to the Sloan Digital Sky Survey (SDSS) 
Data Release 9 \citep{ahn12} astrometric frame by
correlating the positions of X-ray sources with SDSS sources. 
An image
was formed in the full 0.2-10 keV energy band from events of ASCA grades 0, 2, 3, 4 and 6. Source detection
was carried out with the {\sc ciao} task {\sc wavdetect}, with wavelet scale sizes of 1, 2, 4 and 8 pixels and a false-positive probability threshold of $10^{-5}$.

\subsection{\em XMM-Newton}
\label{sec:xmmobs}

\ulas\ was observed over three {\em XMM-Newton} orbits between 2012
May 23 and June 21 for a total observing time of
331~ks. 
The European Photon Imaging Cameras (EPICs) were operated in full-frame mode, 
with thin filters. 
EPIC data were processed
using the {\em XMM-Newton} {\sc science analysis software} ({\sc sas}) version
12.0.1 \citep{gabriel12}.  
Times of high particle background were excluded by inspection
of lightcurves in the 5-12 keV energy range, leading to total exposure
times of 226, 223 and 174~ks in the MOS1, MOS2 and pn cameras
respectively.  Images were constructed in four bands: 0.2-0.5~keV,
0.5-2~keV, 2-5~keV and 5-10~keV. Background images were produced
following the procedure described in \citet{loaring05}. The images
were then source-searched in the four energy bands simultaneously
using the standard SAS tasks {\sc eboxdetect} and {\sc emldetect},
again following the procedure described in \citet{loaring05}. 

We extracted a spectrum of \ulas\ following the procedure described in
\citet{page06}, in this case extracting source counts from an
11~arcsec radius region around the target in each detector. 
This source-extraction region includes 60--70 per cent of the photons for a point source such as \ulas, the precise fraction depending on photon energy and EPIC camera. The small aperture is chosen to minimise the background contribution to the spectrum of this very faint source. The enclosed energy fraction is taken into account in the generation of the response files by the standard {\sc sas} task {\sc arfgen}.
Event
patterns 0-12 were included in the MOS cameras, while for the pn
camera we used patterns 0-4 above 0.4 keV and only pattern 0 between
0.2 and 0.4 keV. Channels containing strong instrumental emission
lines were excluded. The spectra of the target from the different
observations and different EPIC cameras were then combined to form a
single spectrum, and the corresponding response matrices and
background spectra were combined in an appropriate fashion to form a
single response matrix and a single background spectrum, following the
method described in Appendix A of \citet{page03}. Finally, the spectrum was 
grouped to a minimum of 20 counts per bin.

\section{Results}
\label{sec:results}

\begin{figure}
\includegraphics[width=84mm,angle=0]{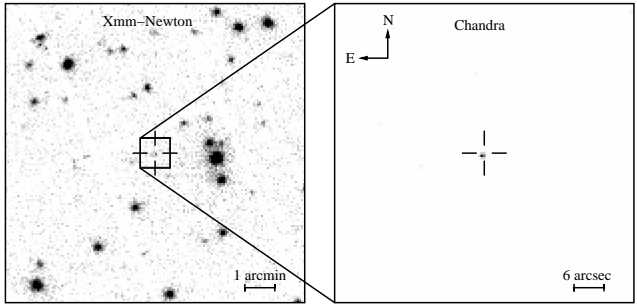}
\caption{The left panel shows the {\em XMM-Newton} EPIC image centred
  on \ulas\ in 0.2-2.0~keV, formed by summing the 0.2-0.5~keV and
  0.5-2.0~keV images described in Section~\ref{sec:xmmobs}. The right
  panel shows the central arcminute of the {\em Chandra} image, after
  adaptive smoothing with the {\sc ciao} task {\sc dmimgadapt}. In
  both panels the cross hairs mark the near-IR position of \ulas\ from
  \citet{mortlock11}}
\label{fig:images}
\end{figure}

In the {\sc wavdetect} source search of the  0.2--10~keV  {\em
  Chandra} ACIS image a point-like source is found within 0.5 arcsec
of \ulas\ at equatorial coordinates 11 20 01.50 +06 41 23.9 (see
Fig.~\ref{fig:images}).  The source has a 1$\sigma$ position
uncertainty of 0.4 arcsec and so its position is consistent with that
of \ulas.  
  The source is formed of 7 net counts (to 1 background
  count), according to the {\sc wavdetect} algorithm, implying
  7.0$^{+4.0}_{-2.8}$ net source counts if we adopt the 68 per cent Poisson
  confidence limits described in \citet{gehrels86}. We have verified
  the source-count measurement by manual inspection of the ACIS image,
  finding 8 counts in the image within a 1 arcsec radius aperture
  around the source, and an average background level in the
  surrounding image that corresponds to 0.7 counts in an aperture of
  this size.
No other X-ray sources are found within 30 arcsec of
\ulas, so we do not expect issues with source blending in the larger
point-spread function of {\em XMM-Newton}.

In the {\em XMM-Newton} EPIC images, an X-ray source is again found with a
position consistent with \ulas: 1.7 arcsec distance, with a
1\,$\sigma$ position uncertainty of 1.3 arcsec (see Fig.~\ref{fig:images}). 
The source is detected with
114 net source counts, 
in the full 0.2-10 keV energy range
with
a false source probability of $1.6\times 10^{-10}$,
equivalent to 6.4\,$\sigma$. It is detected individually in the
0.2-0.5~keV and 0.5-2.0~keV bands with fluxes of $6.2\pm 1.7\times
10^{-16}$~erg~s$^{-1}$~cm$^{-2}$ and $5.7\pm 1.2\times
10^{-16}$~erg~s$^{-1}$~cm$^{-2}$ respectively. It is not detected
above 2 keV, and 3 sigma upper limits are obtained of $4.1\times
10^{-16}$~erg~s$^{-1}$~cm$^{-2}$ for the 2-5 keV flux and $4.2\times
10^{-16}$~erg~s$^{-1}$~cm$^{-2}$ for the 5-10~keV flux. 

\begin{figure}
\includegraphics[width=84mm,angle=0]{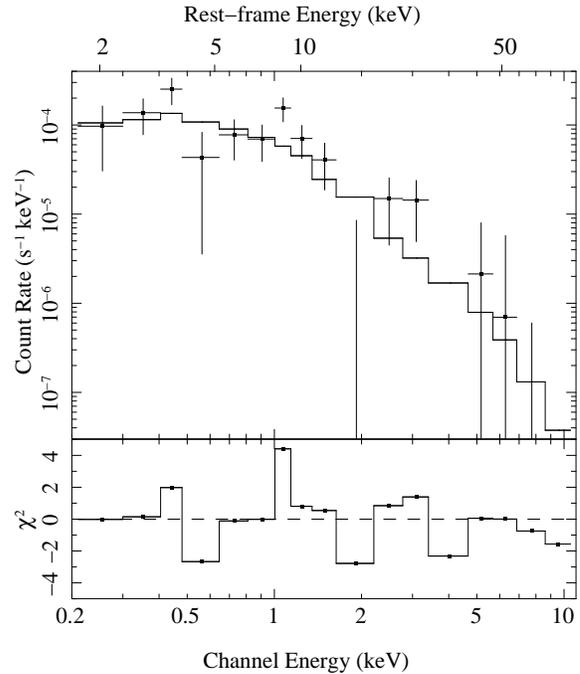}
\caption{{\em XMM-Newton} EPIC spectrum of \ulas. The upper panel
  shows the observed counts spectrum (datapoints) together with the
  best fit power-law spectral model, with $\alpha =1.64$ and Galactic
  $N_{H}=5.07\times 10^{20}$~cm$^{-2}$. The lower panel shows the 
  contribution of each bin to the total $\chi^{2}$ multiplied by the 
  sign of (data $-$ model).}
\label{fig:spectrum}
\end{figure}

The X-ray spectrum of \ulas\ obtained from the EPIC data is shown in
Fig.~\ref{fig:spectrum}. The spectrum was fitted with a power law model,
with fixed Galactic absorption.  The best fit spectral index is
$\alpha$=1.64$^{+0.37}_{-0.33}$. The fit is acceptable, with a
$\chi^{2}$ of 20.4 for 15 degrees of freedom\footnote{Given the
relatively small number of counts, we have also performed the fitting
using the $C$-statistic \citep{cash79} instead of $\chi^{2}$, and
obtain a very similar best-fit slope of $\alpha =
1.58^{+0.30}_{-0.35}$.}.
 The contribution of each channel to
the $\chi^2$
is shown in the bottom panel of Fig.~\ref{fig:spectrum}. There are no
deviations from the best-fit model which are individually
significant, and there is no systematic shape to the $\chi^{2}$
contributions which might justify a more complex model. 
The rest-frame 2-10 keV luminosity implied by the 
fit is $4.7\pm0.9\times 10^{44}$~erg~s$^{-1}$.

A common measure of the UV to X-ray spectral shape in AGN is the power
law slope $\alpha_{OX}$ that would connect the flux densities at
2,500~\AA\ and 2 keV in the rest frame of the source. For \ulas,
restframe 2,500~\AA\ falls at 20,213~\AA, in the near-IR in the
observed frame. We measure the flux density at this wavelength from
the near-IR spectrum presented in \citet{mortlock11} to be $3.5\pm0.4\times
10^{-18}$~erg~s$^{-1}$~cm$^{-2}$~\AA$^{-1}$. 
An energy of 2~keV in the restframe of \ulas\ falls at
0.247 keV in the observed EPIC spectrum. We obtain the 0.247 keV flux
and the uncertainty in this value directly from the power-law fit to
the EPIC spectrum. 
Assuming that the rest-frame ultraviolet emission was the same at 
the time of the {\em XMM-Newton} observation as it was at the time 
of the near-IR observations,
 we find $\alpha_{OX}=1.76\pm0.07$, where the uncertainty is 
dominated by
the uncertainty on the 0.247 keV flux.

To compare the X-ray flux between the {\em Chandra} and {\em
  XMM-Newton} observations, we note that the X-ray flux and spectrum
of \ulas\ are characterised much better in the {\em XMM-Newton}
observation than in the {\em Chandra} observation. An appropriate way
to perform the comparison is therefore to use the parameters from the
spectral fit to the {\em XMM-Newton} spectrum to estimate the number
of counts that would be expected in the {\em Chandra} observation, for
comparision to the observed number of counts. For this, we use version
4.6b of the {\sc portable, interactive multi-mission simulator} \citep[{\sc
  pimms}; ][]{mukai93}. The best-fit parameters from the XMM-Newton
spectrum, correspond to an expected 1.9 source counts from \ulas\ in
the 0.2-10~keV band {\em Chandra} S3 image, and the uncertainties in 
the {\em XMM-Newton} spectrum
translate to an uncertainty of 20 per cent in the predicted number of
{\em Chandra} counts.

Including the background level in the
{\em Chandra} image (1 count in the {\sc wavdetect} cell), we would
predict a total of 2.9$\pm0.4$ counts at the position of \ulas\
in the {\em Chandra} image, compared to a total of 8 counts
observed. Assuming Poisson statistics, the probability of
observing 8 or more counts for an expectation of 2.9 counts is only
0.01, providing strong evidence that \ulas\ has decreased
in X-ray flux in the 15 months  
between the {\em Chandra} and {\em XMM-Newton}
observations ($<2$ months in the rest frame of the quasar). 
Comparable variability, while not the norm, has been observed before in high redshift quasars \citep[e.g. ][]{shemmer05}.
Assuming the best-fit {\em XMM-Newton} spectrum, the count rate measured in the {\em Chandra} image corresponds to a rest-frame 2-10~keV luminosity of $1.8^{+1.0}_{-0.7} \times 10^{45}$erg~s$^{-1}$, and if we assume that the rest-frame ultraviolet flux is the same at the times of the near-IR and X-ray observations we obtain $\alpha_{OX}=1.54^{+0.09}_{-0.08}$ at the time of the {\em Chandra} observation.

We also searched for variability during the month in which the {\em
  XMM-Newton} observations were carried out. We measured the count
rate in the 0.2-2.0 keV range, in the three {\em XMM-Newton}
observations separately. The three count rates are all within
1~$\sigma$ of the mean count rate; hence there is no evidence for 
variability during the {\em XMM-Newton} observation period.

Finally, we have examined the optical counterparts to the X-ray
sources found with {\em XMM-Newton} and {\em Chandra} 
within a 200 arcsec radius of
\ulas, where we have deep $z$-band and $Y$-band near-IR imaging, to
see if any are likely to be at the same redshift as the quasar. None
of the candidates brighter than $Y$=20.5 have sufficiently red $z-Y$
colours to indicate a redshift $z\sim7$, and so there is no evidence for
X-ray sources which are in a common large-scale
structure with \ulas. 
The X-ray sources within 200 arcseconds of 
\ulas\ are likely to be unrelated AGN at lower redshifts.

\section{Discussion}
\label{sec:discussion}

There is a well-documented linear correlation between $\alpha_{OX}$ and log
UV luminosity, such that the most UV-luminous quasars have the largest
values of $\alpha_{OX}$, and hence the largest ratios of UV to X-ray
luminosity \citep[e.g. ][]{strateva05, steffen06, just07, vasudevan09,
  wu12}.  According to the relation given by \citet{just07}, and
translating the monochromatic 2,500~\AA\ luminosity of \ulas\
($3.5\times 10^{31}$~erg~s$^{-1}$~Hz$^{-1}$) to the cosmology assumed
by \citet{just07}, we would predict a value of
$\alpha_{OX}=1.71\pm0.14$ for \ulas, where the uncertainty represents
the observed dispersion around the relation. The value of
$\alpha_{OX}=1.76\pm0.07$ obtained from the {\em XMM-Newton} observation 
is thus consistent with the value
expected, and $\alpha_{OX}$ is still consistent with the prediction if the 
X-ray flux is increased 
to the level suggested by the {\em Chandra} observation.

On the other hand, the X-ray spectral slope found for \ulas,
$\alpha$=1.64$^{+0.37}_{-0.33}$, is somewhat softer than the mean
found for lower redshift quasars. 
  For optically-selected quasar samples with mean redshifts below 2,
  observed with {\em XMM-Newton}, \citet{young09} and \citet{scott11}
  obtained mean values for $\alpha$ of $\langle \alpha\rangle=0.90$
  and $\langle \alpha\rangle=0.99$ with intrinsic dispersions of
  $\sigma_{\alpha} = 0.40$ and $\sigma_{\alpha} = 0.30$
  respectively. Similarly, studies of X-ray-selected quasar samples
  with mean redshifts below 2 generally find $0.89 < \langle
  \alpha\rangle < 1.00$, with dispersions around the mean ranging
  from $\sigma_{\alpha} = 0.2$ to $\sigma_{\alpha} = 0.36$
  \citep[e.g. ][]{mateos05, page06, mateos10, lanzuisi13}.

Arguably, samples of optically selected quasars form the best comparison
samples for \ulas, given its discovery via rest-frame UV
emission. Furthermore, given its large redshift such that emission
below 2 keV in the rest frame is shifted below the {\em XMM-Newton }
bandpass in the observed frame, spectral indices measured above 2 keV
in the rest frame are the most appropriate for comparison with that of
\ulas. Conforming to these two criteria, for data above 2 keV in the rest frame 
\citet{piconcelli05} find
$\langle \alpha\rangle=0.89$ for radio-quiet PG quasars, while 
\citet{page04} find $\langle
\alpha\rangle=0.90$ for their small sample of luminous,
optically-selected quasars.
 Similar to the comparison
with X-ray selected samples, \ulas\ has a softer X-ray spectrum than
the average for optically selected quasars at lower redshift.

Moving now to high-redshift quasars, 
\citet{grupe06} examined the {\em XMM-Newton} observations of
a sample of 21 quasars at $z>4$. For the 10 radio-quiet objects in
their sample which were detected with enough counts to permit spectral
analysis, they obtained\footnote{The uncertainty given here is the
  error on the mean, whereas \citet{grupe06} give the
  standard deviation.} $\langle
\alpha\rangle=1.19\pm0.16$, and argued that this rather soft slope
indicates that $z>4$ quasars are accreting at a high fraction of the
Eddington rate. On the other hand, \citet{just07} obtained $\langle
\alpha\rangle=0.93\pm0.16$ for a sample of $z>4$ radio quiet quasars
using a combination of {\em XMM-Newton} and {\em Chandra} data, and
\citet{shemmer06} found $\langle \alpha\rangle=0.95^{+0.30}_{-0.26}$
for their sample of 15 $z>5$ radio-quiet quasars using {\em Chandra}
observations. Both of these samples yield X-ray spectral slopes which
are indistinguishable from quasar samples at lower redshift. For the
individual $z=6.3$ radio-quiet quasar SDSS~J1030+0524,
\citet{farrah04} found $\alpha = 1.12\pm0.11$, and argued that neither
this quasar, nor other high-redshift quasars which had been studied at
that time, could be distinguished from quasars at lower redshifts by
their X-ray spectra. \ulas\ however has a softer X-ray spectrum than
is typical for any of these samples.

It is well established \citep{pounds95, leighly99} that X-ray
spectral indices $\alpha > 1$ are associated with large
ratios of $L/L_{\rm Edd}$ where $L$ is bolometric luminosity 
and $L_{\rm Edd}$ is Eddington luminosity. 
Indeed, it has been proposed that $\alpha$ can be used as an
estimator of $L/L_{\rm Edd}$ by using the  
linear correlation between log ($L/L_{\rm Edd}$) and 
$\alpha$ \citep{shemmer08}. Substituting the
measured $\alpha$ for \ulas\ into the expression relating $\alpha$ and
$L/L_{\rm Edd}$ given by Equation 1 of \citet{shemmer08}, or the
equivalent expressions from later works \citep{risaliti09, jin12,
  brightman13} suggests that \ulas\ may be accreting at several times the
Eddington rate.
For example, adopting the expression
from \citet{jin12}, and deriving the overall uncertainty by adding the
measurement error on $\alpha$ to the intrinsic dispersion of $\alpha$
about the relation in quadrature, we obtain
$L/L_{\rm Edd}=5^{+15}_{-4}$.

Since our X-ray observations suggest that $L/L_{\rm Edd}>1$, and
moderate levels of super-Eddington accretion are consistent with the
range of $L/L_{\rm Edd}$ estimated from the UV luminosity /
emission-line-derived black-hole mass of \ulas\ 
\citep[$L/L_{\rm Edd}=1.2^{+0.6}_{-0.5}$][]{mortlock11} we briefly explore how
super-Eddington accretion might impact our understanding of this
object.

First, we consider the unusually small ionized near-zone around \ulas. 
\citet{bolton11} show 
that the quasar  cannot have been shining at its present UV brightness for
significantly more than $10^{7}$ years, which is only one fifth of a 
characteristic e-folding timescale of $5\times 10^{7}$~years for 
Eddington-limited accretion. 
The e-folding timescale is inversely proportional to $L/L_{\rm Edd}$, 
so the discrepancy between the two timescales is reduced by a factor of 2 if 
$L/L_{\rm Edd}$ = 2. 

Second, we examine the implications of super-Eddington accretion for
the constraints on the seed black hole. 
Considering that the bolometric
luminosity of the quasar is rather better determined than its black
hole mass, we can take the black hole mass to scale inversely with
$L/L_{\rm Edd}$; adopting the luminosity and mass estimates 
from \citet{mortlock11}, this implies 
 $M_{BH}= 2.4 (L_{\rm Edd} / L) \times
10^{9}$~M$_{\odot}$. More importantly, the e-folding timescale is
inversely proportional to $L/L_{\rm Edd}$, implying a larger ratio of
final to seed black hole mass for higher $L/L_{\rm Edd}$. For reference, 
we would require seed black holes 
of $5.6 \times 10^{3}$~M$_{\odot}$
and $2.7 \times 10^{4}$~M$_{\odot}$ for formation redshifts of
$z=30$ and $z=20$ respectively if the quasar has had $L/L_{\rm Edd}
= 1$ since the formation of its seed. The assumption that the quasar has 
accreted continuously, and without significant off-periods, i.e. a duty cycle 
close to 1, would be unrealistic for quasars with $z<3$, but is reasonable 
for quasars with $z>4$ \citep{shankar10}. 
To have grown the black hole in
\ulas\ from a 100~M$_{\odot}$ seed would require an average value of
$L/L_{\rm Edd}=1.29$ over the lifetime of the quasar if the seed
formed at $z=30$ or $L/L_{\rm Edd}=1.46$ if the seed formed at $z=20$.
Such values of $L/L_{\rm Edd}$ are compatible with the observational
constraints, hence \ulas\ could have grown from the remnant of a
Population III star, as well as from the remnant of a quasi-star, or
from a collapsed stellar cluster \citep{begelman06, volonteri12}, if it has
maintained such a rate of growth throughout its existence.

\section*{Acknowledgments}
\label{sec:acknowledgments}

This work is based in part on observations obtained with {\em
  XMM-Newton}, an ESA science mission with instruments and
contributions directly funded by ESA Member States and NASA.
  This work is based in part on observations made by the {\em Chandra}
  X-ray Observatory
 and has made use of
software provided by the {\em Chandra} X-ray Center in the application
package CIAO.

\bibliographystyle{mn2e}

\label{lastpage}
\end{document}